# Fabrication of high quality factor lithium niobate double-disk using a femtosecond laser

Zhiwei Fang,[1,3,4] Ni Yao,[2] Min Wang,[1,3] Jintian Lin,[1] Jianhao Zhang,[1,3] Rongbo Wu,[1,3] Lingling Qiao,[1] Wei Fang,[2] Tao Lu,[5,*] and Ya Cheng[1,6,7,†]

[1] State Key Laboratory of High Field Laser Physics, Shanghai Institute of Optics and Fine Mechanics, Chinese Academy of Sciences, Shanghai 201800, China

[2] State Key Laboratory of Modern Optical Instrumentation, College of Optical Science and Engineering, Zhejiang University, Hangzhou 310027, China

[3] University of Chinese Academy of Sciences, Beijing 100049, China

[4] School of Physical Science and Technology, ShanghaiTech University, Shanghai 201210, China

[5] Department of Electrical and Computer Engineering, University of Victoria, Victoria, BC, V8P 5C2, Canada

[6] State Key Laboratory of Precision Spectroscopy, East China Normal University, Shanghai 200062, China

[7] Collaborative Innovation Center of Extreme Optics, Shanxi University, Taiyuan, Shanxi 030006, China

* taolu@ece.uvic.ca

† ya.cheng@siom.ac.cn

## Abstract

We demonstrate fabrication of a high quality factor lithium niobate double-disk whispering-gallery microcavity using femtosecond laser assisted ion beam milling. Using this method, two vertically stacked 30-μm-diameter disks with a 200-nm-gap are fabricated. With our device, an optical quality factor as high as $1.2\times10^5$ is demonstrated. Our approach is scalable to fabricate multiple disks on a single chip.

# 1. INTRODUCTION

High-quality (Q) factor whispering-gallery microcavities (WGMs) have attracted much attention for their broad range of applications ranging from optical signal processing and cavity quantum electrodynamics to biosensing and optomechanics [1-10]. Currently, high-Q WGMs are mostly fabricated with e-beam or photo lithography, thermal reflow as well as mechanical milling and polishing technologies [11-13]. The mechanical approach lacks the potential for monolithic integration of multiple WGMs for on-chip applications while the lithography and reflow ones are limited by materials used in WGMs. Therefore, developing efficient fabrication techniques for producing on-chip crystalline WGMs is still challenging due to the incompatibility of a large number of crystalline materials and optical lithography. Nevertheless, it has recently been shown that femtosecond laser micromachining provides a promising approach to fabricating high-Q WGMs on various materials including glasses [14-16], polymers [17,18], and crystals [19-21]. It is noteworthy that the successful demonstrations of high-Q WGMs on crystalline substrates open the door for miniaturized nonlinear optics applications.

In the past few years, significant nonlinear optical phenomena and efficient electro-optic tuning effects have been experimentally demonstrated in lithium niobate (LN) WGMs [22-26]. In addition, the capability of monolithic integration of LN microresonators with various nanophotonic structures has been reported [25-27]. To date, only single-layered LN micro-disks have been fabricated using the LN thin film substrate. It is known that silica double-disk is one of the unique structures which displays strong optomechanical effects due to the large optical gradient force provided by the strong interaction of optical fields between the top and bottom disks [28-32]. Here, we fabricate double-disks on LN platform by utilizing a femtosecond laser in combination with focused ion beam as the LN crystal has advantageous nonlinear optical, mechanical and electro-optical properties compared to SiO$_2$. The unique physical properties of LN can influence the optomechanical responses in WGMs and in turn provide opportunities to new findings and applications.

## 2. EXPERIMENT

We designed the double-layer X-cut LN thin film substrate as illustrated in Fig. 1(a). The top and bottom LN thin films of 300 nm in thickness, respectively, are separated by a thin layer of SiO$_2$ with a thickness of 200 nm. The double-layer LN thin film is bonded a 2-μm-thick SiO$_2$ substrate, which is bonded to the 500-μm-thick LN substrate. Following our design, the wafer was produced by NANOLN, Jinan Jingzheng Electronics Co., Ltd. A Ti: sapphire femtosecond laser source (Coherent, Inc., center wavelength: 800 nm, pulse width: 40 fs, repetition rate: 1 kHz) was used for fabricating the on-chip double-disk LN microresonator. A variable neutral density filter was used to tune the average power of the laser beam. In the femtosecond laser direct writing process, an objective lens (100× / NA 0.80) was used to focus the beam down to a ~1 μm-diameter focal spot. The sample could be arbitrarily translated in 3D space at a resolution of 1 μm using a PC-controlled XYZ stage combined with a nano-positioning stage. A charged coupled device (CCD) connected to the computer was installed to monitor the fabrication process in real time.

The procedures of fabricating the LN double-disk WGM are schematically illustrated in Fig. 1. First, the LN substrate was immersed in water and ablated with tightly focused femtosecond laser pulses, as shown in Fig. 1(a). The ablation in water can help reduce the debris and cracks in the fabricated structure. The height of the cylindrical microstructure patterned with femtosecond laser ablation is ~5 μm, as shown in Fig. 1(b). The femtosecond laser fabrication took about 1 hour. Next, the periphery of the LN double-disk WGM were smoothed using focused ion beam (FIB) milling, as illustrated in Fig. 1(c). In the FIB milling, a 30-kV ion beam with a beam current of 1 nA was used. The FIB milling was completed in 10 min. Finally, chemical wet etching, which selectively removes the $SiO_2$ layers underneath the LN thin films to form free-standing double LN micro-disks, was performed in a solution of 2% hydrofluoric (HF) for 8 minutes, as shown in Fig. 1(e). The $SiO_2$ layer was partially preserved to support the double-disk LN microresonator. The diameter of the LN micro-disk is 30 μm. It took about 1.5 hrs in total to produce the LN double-disk WGM.

## 3. Measurement of quality factor and analysis of WGM modes

The optical micrograph in Fig. 2(a) shows the top-view image of the fabricated LN double-disk WGM. The side-view image obtained with the scanning electron micrograph (SEM) is presented in Fig. 2(b), which reveals the detailed geometry of the fabricated LN double-disk. From the right inset in Fig. 2(b), we determine that the thicknesses of top and bottom LN disks are 291 nm and 312 nm, respectively. An air gap of 138 nm is created between the top and bottom disks after the silica layer is partially removed by the chemical etching. In particular, from the left inset in Fig. 2(b), the sidewall of the double-disk displays a tilt angle of 26° with respect to the axis perpendicular to the double-disk plane (see, Fig. 2(b)). This slanting sidewall is caused by the conical ion beam used in the milling process. The surface appears smooth under the SEM examination, which ensures a high-Q factor of the fabricated double-disk as we have demonstrated before [21].

To measure the Q factor of the fabricated LN double-disk WGM, we used a system as shown in Fig. 3. Here, a narrow-bandwidth continuous-wave tunable diode laser (New Focus, Model 688-LN) was used as the light source. The wavelength-tunable light was first coupled into and then extracted from the micro-disk through a tapered optical fiber. A transient photo detector (Lafayette, Model 4650) was connected to the output of the tapered fiber for measuring the transmission spectrum. The WGM modes of quasi-TE polarization states were selectively excited using an in-line fiber polarization controller.

Figure 4(a) shows a recorded transmission spectrum at 1573.22 nm (i.e., the black curve), from which an overall optical Q-factor of $1.2 \times 10^5$ can be determined with Lorentz fitting curve in red. The result clearly shows that our LN double-disk can have a high Q factor on the same level of that of single-disk LN microresonator. Further, we attempted to identify our measured modes with numerical simulations. Here, a full vector WGM mode solver was implemented using COMSOL 2-D axisymmetric, wave optics frequency domain electromagnetics module. In this simulation, we assume the refractive index of LN to be $2.2111 + 2.467 \times 10^{-8} i$ [33]. The edge of the double disk is centered in an 8-μm wide and 3-μm high computation window. To minimize the spurious reflection, 1.55-μm thick perfect match layers were added at the edges of the window. The inset in Fig. 4(a) shows the simulated WGM profile in the double disk, which corresponds to the fundamental quasi-TE mode at an azimuthal order of 102 at a resonance wavelength of 1572.16 nm. The resonant wavelengths obtained from experimental measurement and simulation nicely agree with each other, indicating that the measured mode is highly likely to be the mode shown in the inset. The slight difference in resonance wavelengths between the experiment and simulation can be attributed to the perturbation from the fiber taper, which can be modelled through perturbation [34] or mode matching methods [35,36]. We note that due to the existence of high-order modes and lack of full geometric information of the double-disk microcavity, complete determination of the modes by measurements requires more sophisticated techniques, which will be practiced in the future. The simulation also indicated an optical Q of $4.8\times10^7$ after only considering the absorption loss, corresponding to the ultimate Q value of such devices after optimizing our fabrication processes (i.e., by suppressing the scattering loss at the sidewalls as much as possible). Further, the plot of major transverse field in the left inset of the plot confirms that the mode is bonded in nature.

Furthermore, Fig. 4(b) shows another measured Q-factor of $3.6\times10^4$ at 1566.13 nm, which is derived from Lorentz fitting as well. According to the simulation, the experimentally observed mode in Fig. 4(b) may be attributed to a high order transverse mode at an azimuthal order of 96 with a resonance wavelength of 1565.91 nm. In contrast to the result reported in ref. [29], our simulations show that the 64° wedge angle leads to degradation of Q for anti-bond modes by an order of magnitude. Therefore, we concluded that the anti-bond mode was unlikely to be excited in our LN double-disk structure.

## 4. CONCLUSION

To conclude, we have demonstrated fabrication of LN double-disks of Q factors on the order of $10^5$. Our technique combines femtosecond laser direct writing and FIB milling, thereby providing both high fabrication efficiency and ultra-high fabrication resolution. We envisage that such device is an excellent platform or the study of cavity QED where the LN provides another dimension of control to the corresponding experiments. In future researches, we will exploit the unique electro-optical and nonlinear optical properties of such device to provide the tunability for investigating cavity optomechanics and nonlinear optics [26].

## 5. ACKNOWLEDGMENTS

National Basic Research Program of China (Program 973) (2014CB921300), Natural National Science Foundation of China (NSFC) (61590934, 61505231, 61405220, 61327902, 61590934), the Fundamental Research Funds for the Central Universities and the Open Fund of the State Key Laboratory on Integrated Optoelectronics (IOSKL2015KF34). T.L. would like to acknowledge Natural Sciences and Engineering Research Council of Canada (NSERC) Discovery (RGPIN-2015-06515) and CMC Microsystems.

**Figure Captions:**

Fig. 1: The processing flow of fabricating an on-chip LN double-disk WGM. (a) Fabrication of the LN double-disk WGM using femtosecond laser microfabrication. (b) The structure obtained after the laser fabrication. (c) Focused ion beam (FIB) milling to smooth the periphery of the LN double-disk WGM. (d) The structure obtained after the FIB milling. (e) Chemical wet etching of the sample undergone the FIB milling to form the freestanding LN double-disk WGM. (f) The structure obtained after the chemical wet etching.

Fig. 2: (a) Top view optical micrograph and (b) side view SEM image of the 30-μm LN double-disk WGM fabricated with femtosecond laser micromachining combined with FIB milling.

Fig. 3: Experimental setup for recording the transmission spectra of the LN double-disk WGM.

Fig. 4(a) Q-factors of (a) $1.2\times10^5$ measured at 1573.22 nm and (b) $3.6 \times 10^4$ measured at 1566.13 nm. The corresponding mode profiles revealed by the simulations are shown in the insets.

Fig. 1

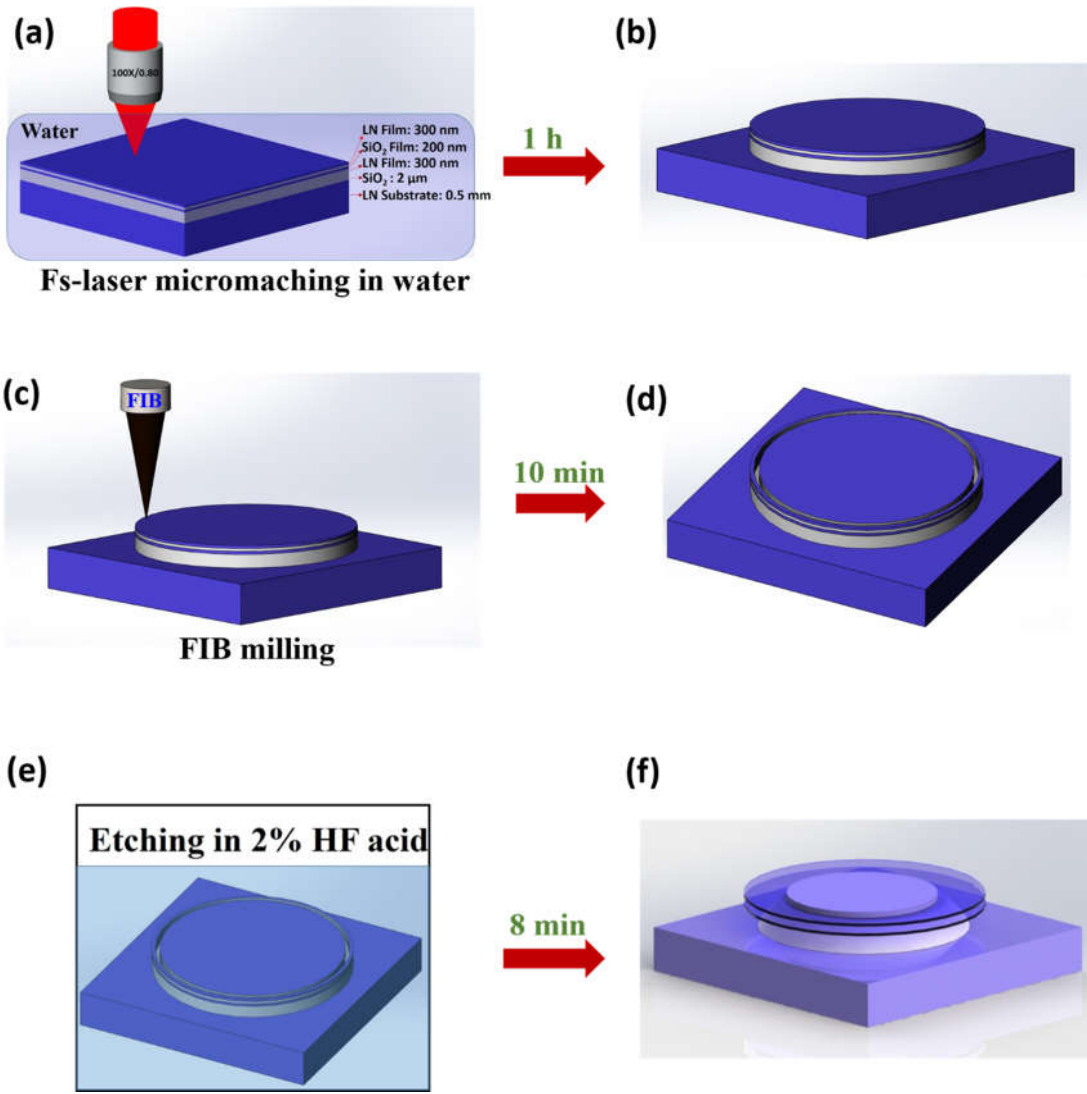

Fig. 2

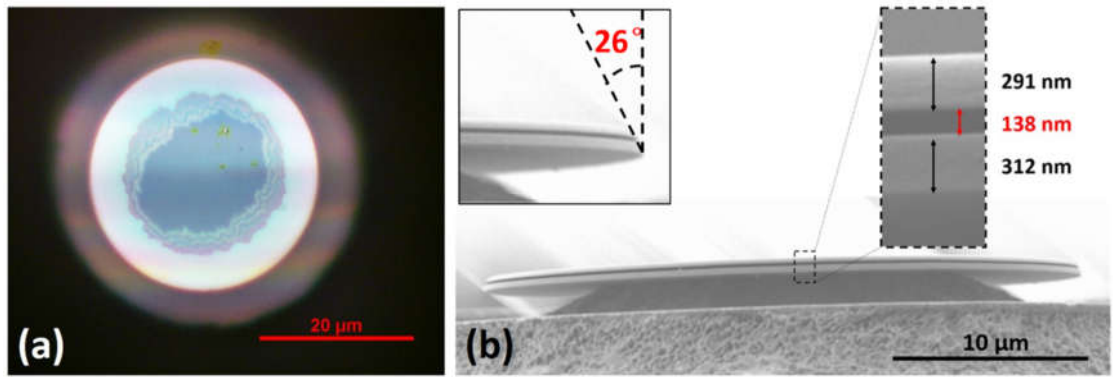

Fig. 3

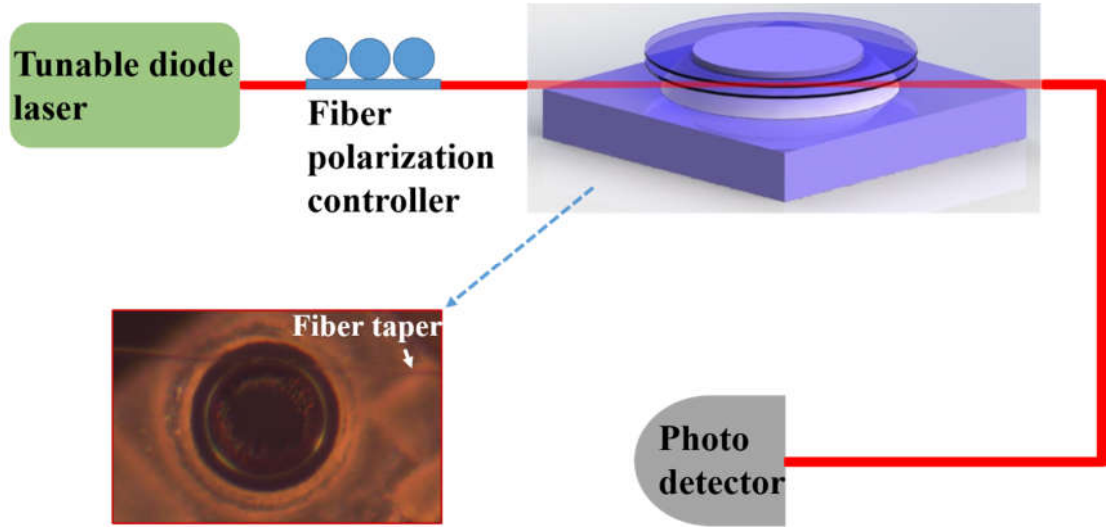

Fig. 4

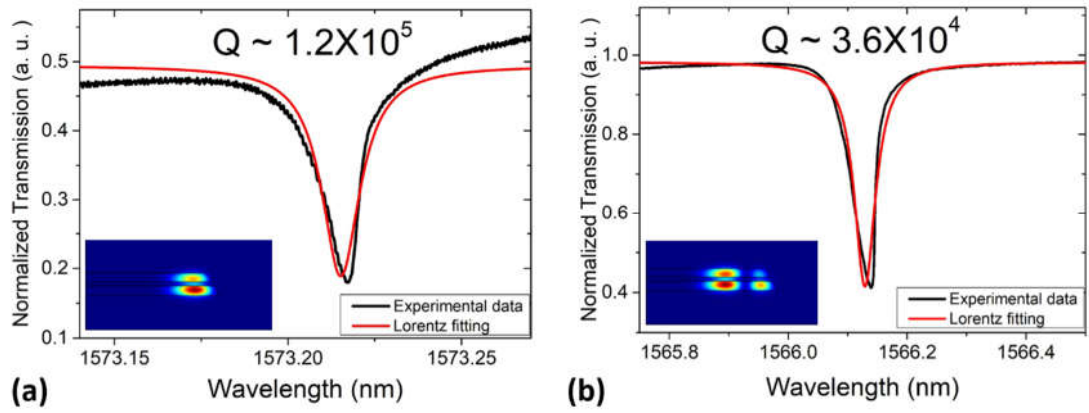